\documentclass[aps,pra,twocolumn, showpacks]{revtex4-1}

\usepackage{graphicx,epstopdf}
\usepackage{amsmath}
\usepackage{amssymb}
\usepackage{braket}
\usepackage{color}
\usepackage{float}
\usepackage{epstopdf}

\begin{document}

\renewcommand*{\DefineNamedColor}[4]{%
   \textcolor[named]{#2}{\rule{7mm}{7mm}}\quad
  \texttt{#2}\strut\\}

\definecolor{red}{rgb}{1,0,0}

\title{Survival of time-evolved correlations depends on whether quenching is across critical point in $XY$ spin chain}

\author{Utkarsh Mishra\(^{1}\), Debraj Rakshit\(^{1}\), and R. Prabhu\(^{1,2}\)}

\affiliation{\(^1\)Harish-Chandra Research Institute, Chhatnag Road, Jhunsi, Allahabad 211019, India\\
\(^2\)Department of Physics, Ben-Gurion University of the Negev, Beersheba 8410501, Israel​}

\begin{abstract}
The time-dynamics of quantum correlations in the quantum transverse anisotropic $XY$ spin chain of infinite length is studied at zero as well as finite temperatures. The evolution occurs due to the instantaneous quenching of the coupling constant between the nearest-neighbor spins of the model, which is either performed within the same phase or across the quantum phase transition point connecting the order-disorder phases of the model. We characterize the time-evolved quantum correlations,  entanglement and quantum discord, which exhibit varying behavior depending on the initial state and the quenching scheme. We show that the system is endowed with enhanced bipartite quantum correlations compared to that of the initial state, when quenched from ordered to the deep disordered phase. However, bipartite quantum correlations are almost washed out when the system is quenched from disordered to the ordered phase with the initial state being at the zero-temperature. Moreover, we identify the condition for the occurrence of enhanced bipartite correlations when the system is  quenched within the same phase. Finally, we investigate the bipartite quantum correlations when the initial state is a thermal equilibrium state with finite temperature which reveals the effects of thermal fluctuation on the phenomena observed at zero-temperature. 
\end{abstract}

\maketitle

\section{Introduction}
Quantum correlations \cite{HHHH, MODIetal} are proven to be useful in realizing many quantum information protocols such as quantum dense coding \cite{DC}, quantum teleportation \cite{Teleportation},  quantum cryptography \cite{cryptography}, and one-way quantum computation \cite{computation}. To quantify the amount of quantum correlation present in a quantum state, measures of two different paradigms, viz., that of entanglement-separability and of information-theoretic, are considered. For example, concurrence \cite{concurrence}  and quantum discord \cite{discord}, belonging to entanglement-separability and information theoretic paradigms, respectively, are in general easy to compute for arbitrary two-qubit systems. It has been argued that quantum discord promises to capture quantum correlations beyond entanglement \cite{discord,MODIetal}. Examples of the physical systems in which quantum correlations can be realized include  photons \cite{photon}, ion-traps \cite{ion-trap}, optical lattices \cite{opt-lattices}, superconductors \cite{supercqubit}, nuclear magnetic resonance \cite{nmr}, etc. Quantum correlation measures, from both paradigms, have been used to detect and study cooperative phenomena like quantum phase transitions \cite{EntMB, DisMB,spinchain_qic}. 
Investigation of many-body systems in the non-equilibrium regime from a quantum information perspective \cite{computation,non-eqi-qic} have gained interest in recent years. 
In this respect, dynamical evolution of a many-body quantum system under sudden quench have been studied in several models \cite{quenched_dynamics} (see \cite{fazio} for a situation where an external environment is present). Note that, experimentally realizable physical systems, e.g., ultracold atoms in an optical lattice \cite{GREINER}, can be prepared in a controlled way, making them effectively isolated from the environment \cite{zollar}. This, in turn, has generated a lot of interest in recent times in the nonequilibrium dynamics of otherwise closed quantum systems due to external disturbance, both experimentally \cite{close_qs_ex} and theoretically \cite{close_qs_th_rev,close_qs_th, DPT}.



Quenching, which involves sudden change in certain  parameters of the system, has attracted interest due to their feasibility in experiments using, for example, cold atomic gases \cite{ simulation_dyn,sudden_q_1, sudden_q_2} (for a review see \cite{close_qs_th_rev}). Spin magnetic systems subjected to sudden quenching have been shown to exhibit non-intuitive behavior, e.g., revival and collapse phenomena of the nearest-neighbor entanglement in the quantum $XY$ model \cite{collapse_revival,DPT}. Ergodicity of various quantum correlations, viz. entanglement and quantum discord have been studied by perturbing system parameters of the quantum $XY$ and $XYZ$ models in low dimensions \cite{ergo_group, sabre} (cf. \cite{barouch}). 
The dynamical evolution of entanglement spectrum at zero temperature has been studied in closed quantum many-body systems crossing a quantum phase transition \cite{spinchain_qic, cross_qpt_1}.

In this paper, we investigate the dynamics of bipartite quantum correlations of the evolved state, starting from the zero-temperature state of the $XY$ spin chain of infinite size,  after a sudden change in the nearest-neighbor coupling strength from some initial value to a final value. 
In the evolution process, the final Hamiltonian is  termed as the driving Hamiltonian. We characterize the quantum correlations for both finite time and large time limits in this model.  
In particular, for varied choices of initial states we study the dependence of the dynamical behavior of quantum correlations, viz. entanglement and quantum discord, on the driving Hamiltonian. Our results show that survival of any finite amount quantum correlation present in a initial state followed by a quench distinctly depends on the driving Hamitonian and on whether quenching is across critical point in $XY$ spin chain.

Specifically, we show that if the coupling constant of the driving Hamiltonian is fixed at a value that corresponds to the disordered phase, then the entanglement remains non-zero, irrespective of the choice of  the initial state. In fact, the bipartite entanglement at large time is enhanced significantly compared to that of the initial state, if the initial state corresponds to the ordered phase and the system is quenched into the disordered phase. On the other hand, if the coupling constant of the driving Hamiltonian is fixed at a value that belongs to the ordered phase of the corresponding Hamiltonian,  then the dynamics show rich features. Specifically, let us mention two important scenarios. First, the quenching is performed across the quantum critical point with the initial and the final values of the interaction strength chosen from two sides of the quantum phase transition point, i.e., the initial state belongs to the disordered phase. In this case, the large time entanglement of the evolved state always vanishes irrespective of the choice of the initial state with finite entanglement. Second, the initial and the final values of the interaction strength are chosen from the same phase, i.e., the ordered phase. In this case, entanglement always survives.  Additionally, we identify quenching strategies via which the long time evolved entanglement of the system is enhanced compared to that of the initial state. We also perform analogous investigations for the quantum discord. Further, we extend our analysis to finite temperature of the initial thermal equilibrium  state. In general, we find that, as it may be expected, quantum discord is more robust against thermal fluctuation in comparison to entanglement.

The paper is organized as follows. In Sec. \ref{sec:qc}, we define the bipartite quantum correlation measures, concurrence and quantum discord, belonging to entanglement separability paradigm and information-theoretic one, respectively. In Sec. \ref{sec:model}, we describe the one-dimensional infinite quantum $XY$ spin model with transverse magnetic field and describe the quenching schemes employed in this work. A brief description of the methodology required  for calculating the exact analytical expressions of the single- and two-body correlators for the system in the thermodynamic limit is provided in Appendix \ref{appendix_sketch}. In Sec. \ref{sec:inst_quench}, we study the behavior of quantum correlations of the evolved system at finite time scales and zero temperature, for a fixed driving Hamiltonian and different choices of initial states. In particular, we study the cases when  the system is quenched within the same phase or across the quantum critical point. In Sec. \ref{sec:large_time}, we discuss the behavior of quantum correlations, both concurrence and quantum discord for the spin model considered here, at infinite time. We also study quantum correlations in such time evolved states at finite temperatures. Finally, we conclude in Sec. \ref{sec:conc}.

\section{Quantum correlation measures}
\label{sec:qc}

In this section,  we briefly introduce the bipartite quantum correlation measures, which are employed in the later parts of this paper.

\noindent\textit{Concurrence}: For any arbitrary two-qubit quantum state, $\rho_{AB}$, concurrence \cite{concurrence} is given by 
$C(\rho_{AB})=\max\{0,\lambda_1-\lambda_2-\lambda_3-\lambda_4\}$,
where $\lambda$'s are square roots of the eigenvalues of $\widetilde {\rho}_{AB}\rho_{AB}$ in descending order with $\widetilde \rho_{AB} = (\sigma_{y}\otimes \sigma_{y})\rho_{AB}^{*} (\sigma_{y}\otimes \sigma_{y})$. The maximum is taken to ensure that the concurrence is set to zero value for separable states. If $C(\rho_{AB})=1$, the state $\rho_{AB}$ is said to be maximally entangled.

\noindent\textit{Quantum Discord}: For an arbitrary two-qubit quantum state, $\rho_{AB}$, quantum discord ($D$) \cite{discord} is defined as the difference of two inequivalent quantum analogues of the equivalent classical mutual information relations. 

The first one is the total correlation, which is quantified as the difference between the sum of von Neumann entropy of the individual subsystems, $S(\rho_{A})$, and $S(\rho_{B})$, and the total system $S(\rho_{AB})$ with $S(\varrho)=-\mbox{Tr}(\varrho \log\varrho)$.  The total correlation is given by
${\cal I} =S(\rho_{A})+S(\rho_{B})-S(\rho_{AB})$,
where $S(\varrho)=-\mbox{Tr}(\varrho \log\varrho)$. The total correlations can be seen as the amount of information shared by the two parties of  quantum state $\rho_{AB}$. 

Another definition for the quantum version classical mutual information is given by
 ${\cal J}=S(\rho_{A})-S(\rho_{A|B})$,
where $S(\rho_{A|B}) = \min_{\{B_i\}}\sum_{i}p_{i}S(\rho_{A|i})$ with the measurement operators, $\{B_{i}\}$, being rank-1 projective operators and $p_{i}$s being the probabilities obtained after the measurements on subsystem $B$. The measured state and the probability of the output state are given by
$\rho_{A|i}=\frac1p_{i} \mbox{Tr}_{B}[( I_{A}\otimes B_{i})\rho_{AB}(I_{A}\otimes B_{i})]$
and 
$p_i=\mbox{Tr}_{AB}[(I_{A}\otimes B_{i})\rho_{AB}(I_{A}\otimes B_{i})]$,
respectively. 

Finally, the quantum discord can be obtained as
$ D={\cal I}-{\cal J}=S(\rho_{B})-S(\rho_{AB})+S(\rho_{A|B})$.
For pure bipartite state, quantum discord reduces to von Neumann entropy of its local density matrix.

\section{The Model}
\label{sec:model}

In this paper, we consider the anisotropic quantum $XY$ spin chain in presence of external transverse magnetic field. The Hamiltonian for this model is given by
\begin{equation}
\label{eq:HXY}
\textsl{H}= \sum_i \frac{J(t)}{4} \left[ (1+\gamma) \sigma^x_i \sigma^x_{i+1} +  (1-\gamma)  \sigma^y_i \sigma^y_{i+1}\right] - \frac{h}{2}\sum_i \sigma^z_i,
\end{equation}
where \(J(t)\) is the time dependent pairwise coupling strength between the nearest-neighbor spins, $h$ is the external transverse magnetic field, and $\gamma$ is the anisotropy constant. The periodic boundary condition, i.e. $\vec\sigma_{N+1}=\vec\sigma_{1}$, is considered. Note that, in the above Hamiltonian, when \(\gamma = 0\), the system corresponds to the $XX$ model and when \(\gamma = 1\), it corresponds to the Ising model. 

The time-dependent coupling constant between the nearest-neighbor spins in the Hamiltonian is chosen as a step function, which is given by
\begin{equation}
\label{eq:jt}
 J(t)= \left\{
 \begin{array}{cc}
 J_{1}, & t\leq 0  \\
 J_{2}, & t>0.
\end{array}\right.
\end{equation}
The sudden change in the coupling constant, i.e., from  $J_1$ at time $t \leq 0$, when the system is prepared in the canonical equilibrium state, $e^{-\beta H(J_1)}$ with $\beta=1/\kappa_{B}T$, $\kappa_{B}$ and T being Boltzmann constant and temperature respectively, to $J_2$ at time $t>0$, when the system is unitary evolving under the influence of new Hamiltonian, $H(J_{2})$, with coupling constant $J_2$, is termed as quenching.

In order to characterize the bipartite quantum correlations present in the spin system whose Hamiltonian is given in Eq.~(\ref{eq:HXY}), we need to find the two-site density matrices of the time evolved state of the system. The general time-dependent two-site density matrix for the evolved state is given by 
\begin{eqnarray}
\rho_{12}(t) &=& \frac 14\big[I\otimes I + \sum_{i=x,y,z} m_i(\sigma_i \otimes I + I \otimes \sigma_i) \nonumber \\
& & \hspace{7em} + \sum_{i,\, j=x,y,z}t_{ij}(\sigma_i \otimes \sigma_j)\big],
\label{eq:dm}
\end{eqnarray}
where $m_i=\mbox{Tr}[\rho_1 \sigma_i]$ is the single site  magnetization in the $i^{th}$-direction with corresponding single-site density matrix  $\rho_1=\frac 12 (I+\vec m\cdot \vec\sigma)$, and $t_{ij}=\mbox{Tr}[\rho_{12}(\sigma_i \otimes \sigma_j)]$ are the two-site correlators. The $m_{x}$ and $ m_{y}$  are identically zero as discussed in Refs. \cite{ergo_group, barouch}. For the evolved state, by using Wicks's theorem (as in Refs. \cite{barouch,LSM}), one can show that $xz$ and $yz$ correlators vanish for any given time and also the $xy$ correlator becomes zero at large times. 
For the Hamiltonian given in Eq.~(\ref{eq:HXY}), the magnetization, $m_z$, and the two-site diagonal correlations, $t_{ii}$, can be exactly calculated for the one-dimensional infinite $XY$ spin model with external quenched transverse magnetic field \cite{barouch}. The analytical expression for $m_z$ and $t_{ii}$ can also be  obtained analogously if the quenching is considered in the nearest-neighbor couplings strengths. The exact analytical expressions and details of the calculations for this case are sketched in Appendix \ref{appendix_sketch}.

\section{INSTANTANEOUS QUENCHING IN THE INTERACTION STRENGTH}
\label{sec:inst_quench}

We now consider that the system, whose Hamiltonian is given in Eq.~(\ref{eq:HXY}), starts evolving from the initial canonical state at zero temperature due to sudden quenching in the coupling constant, as given in Eq.~(\ref{eq:jt}), and study the behavior of bipartite quantum correlation measures, both concurrence and quantum discord, with respect to the  evolution time. The evolution of the system is initiated at $t=0$ by an instantaneous change in the nearest-neighbor interaction strength from some initial value, $J_1$, to a final value, $J_2$. Throughout the process of time evolution of the system, the external magnetic field is kept unaltered. Hence, we scale the coupling constants $J$ by $J/h$, which is henceforth denoted as $\widetilde{J}$. It is well known that the static $XY$ Hamiltonian undergoes a quantum phase transition from a ``disordered" phase with $\widetilde{J} < 1$ to an ``ordered" phase with $\widetilde{J} > 1$ at the quantum critical point $\widetilde{J}=1$. The system is considered to be in equilibrium at $t=0$ with Hamiltonian $H(\widetilde J_1)$ and starts evolving after $t>0$ with the new driving Hamiltonian $H(\widetilde J_2)$.

The quantum $XY$ spin chain with transverse magnetic field and time-dependent coupling constant described by the Hamiltonian given in Eq.~(\ref{eq:HXY}) is exactly solvable by successive applications of Jordan-Wigner, Fourier, and Bogoliubov transformations (see Appendix \ref{appendix_sketch}). The two-site density matrices (see Eq. (\ref{eq:dm})), for both initial and evolved states of this spin chain, can be obtained by using the analytical expressions of the magnetization and the two-site correlation, which are given in Appendix \ref{appendix_sketch}. The bipartite quantum correlations for the initial and the evolved states can be computed using these two-site density matrices. 

\begin{figure}
\includegraphics[width=0.5\textwidth]{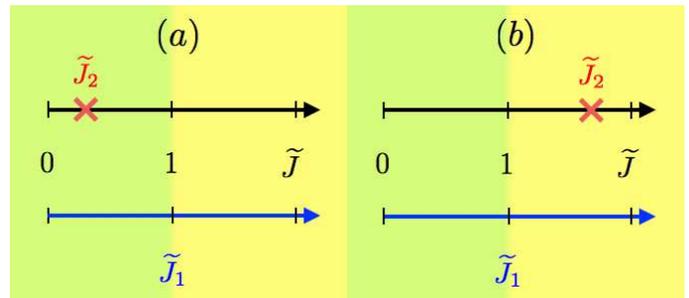}
\caption{(Color online.) The quenching scheme employed in this paper. In case $A$, the final coupling constant is fixed in the disordered phase, i.e.,  $\widetilde{J}_2<1$. In case $B$, the final coupling constant is fixed in the region of ordered phase, i.e.,  $\widetilde{J}_2>1$. In both the cases, the initial coupling constant $\widetilde{J}_1$ is varied from the values ranging from the disorder to the ordered phase. Such situations cover both the scenarios i.e., when $\widetilde{J}_1$ and $\widetilde{J}_2$ are in same phase or $\widetilde{J}_2$ and $\widetilde{J}_2$ are different phases.  All quantities are dimensionless.}
\label{fig:case1}
\end{figure}

During quenching, the choice of the values for $\widetilde{J}_1$ and $\widetilde{J}_2$ can be considered in two different situations: $(i)$ a situation where both $\widetilde{J}_1$ and $\widetilde{J}_2$ are chosen from the same phase (i.e., both $\widetilde{J}_1<1 $ and $\widetilde{J}_{2} < 1$ or both $\widetilde{J}_{1} > 1$ and $\widetilde{J}_{1} > 1$), and $(ii)$ when both of them are  chosen from the different phases. For simplification, we fix $\widetilde{J}_2$ to be in either of the two phases and then probe the behavior of quantum correlations for different initial states by continuously varying the initial values of the coupling constant, $\widetilde{J}_1$, from the disordered phase to the ordered one.  Fig.~\ref{fig:case1} depicts these two situations schematically.
%

\section{Dynamics of Quantum correlations under quenching}

Let us now discuss the behavior of quantum correlation measures, both concurrence and quantum discord, for two different cases depending on the choice of coupling constant $\widetilde J_2$, i.e., Case A:  $\widetilde J_2$ corresponds to the disordered phase (i.e., $\widetilde J_2 < 1$) and Case B:  $\widetilde J_2$ corresponds to the ordered phase (i.e., $\widetilde J_2 > 1$). These cases cover both the scenarios that were mentioned earlier. The characteristics of bipartite quantum correlations in each of these cases are considered separately.

\begin{figure}
\includegraphics[width=0.35\textwidth]{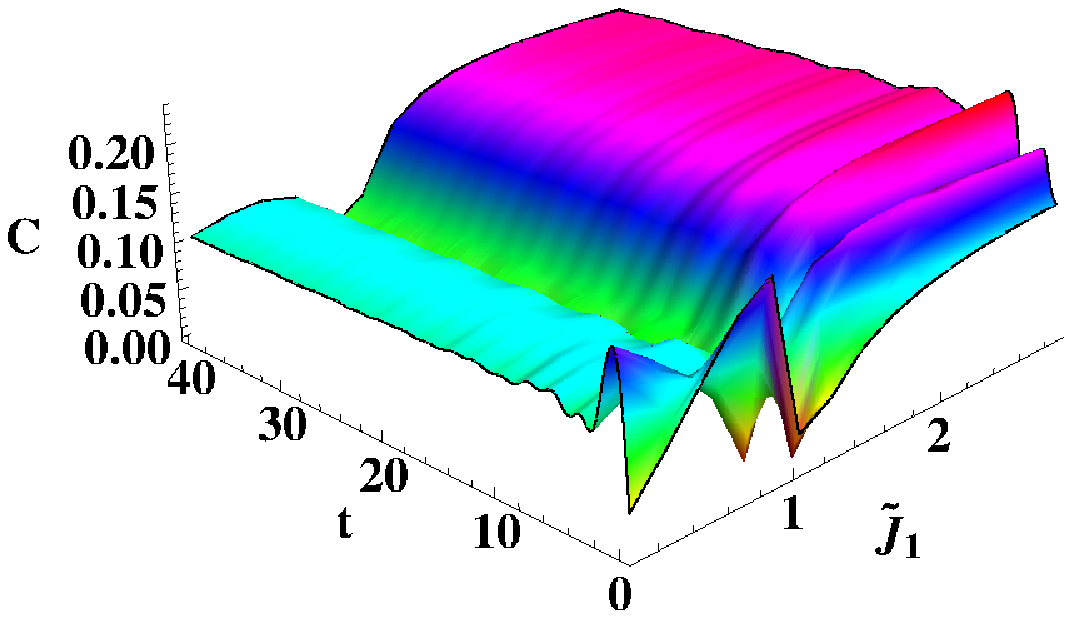}\hspace{1.2cm}
\includegraphics[width=0.35\textwidth]{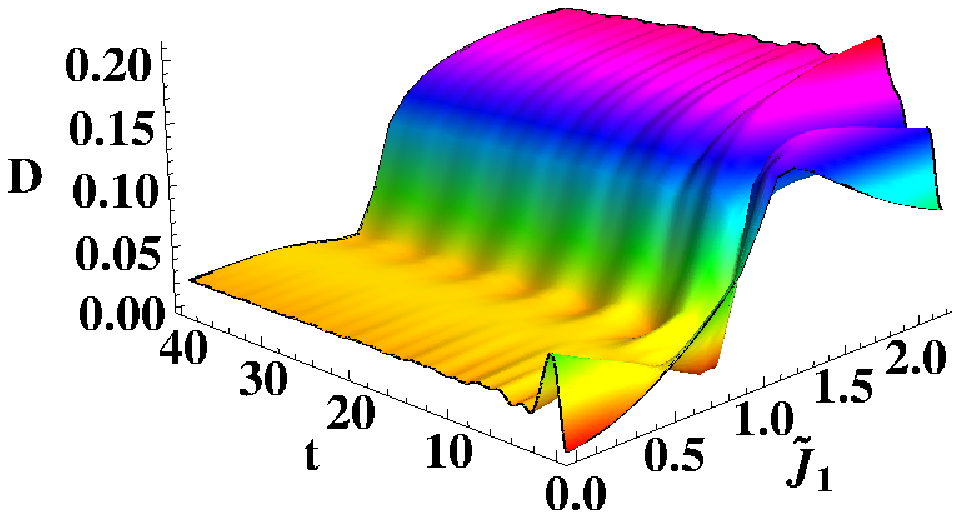}
\caption{(Color online.) Dynamics of nearest neighbor concurrence (top) and quantum discord (below) in the evolved state of quantum $XY$ model against time, $t$, and quenched coupling $\widetilde{J}_1$. The final coupling is fixed at $\widetilde{J}_2=0.5$. We choose  $\gamma=0.5$. The base axes represent dimensionless quantities, while the vertical axis in the top (bottom) panel is in ebits (bits).}
\label{fig:J2_05}
\end{figure}

\subsection*{Case A: $\widetilde J_2$ corresponds to the disordered phase}
\label{subsec:disordered}

Consider first the case when the quenched coupling constant, $\widetilde{J}_2$, at $t>0$ is fixed at a value which corresponds to the disordered phase of the driving Hamiltonian, $H(\widetilde{J}_2)$ (see Fig.~(\ref{fig:case1})). The initial state at $t=0$ corresponds to the zero-temperature state of the system governed by the Hamiltonian $H(\widetilde{J}_1)$. Depending on the choice of $\widetilde{J}_1$, the initial state is  tuned across the disordered and ordered phases. The external applied magnetic field is not altered to keep the uniform scaling of coupling constant throughout the evolution process. 
Using $m_{z}$ and correlators, $t_{ij}$, we evaluate the concurrence and quantum discord for the nearest-neighbor spins, as described in Sec. \ref{sec:qc}. 

In Fig.~\ref{fig:J2_05}, we plot the concurrence (top panel) and quantum discord (bottom panel) for the system under evolution 
with respect to the coupling constant of the initial Hamiltonian $H(\widetilde{J}_1)$ and time ($t$). The transverse magnetic field is kept constant, $h=1$, and anisotropic  constant, $\gamma$, is chosen to be as $\gamma=0.5$. We choose $\widetilde{J}_2=0.5$ and vary both $\widetilde{J}_1$ and time from zero to some higher values. We observe that the behavior of quantum correlations is qualitatively similar for any choice of $\widetilde{J}_2$,  provided $\widetilde{J}_2$ is less than unity. From Fig.~\ref{fig:J2_05}, it is clear that the behavior of quantum correlations can be divided into three different regions in the $\widetilde{J}_1-t$ plane: Region 1 with $\widetilde{J}_1<1$, region 2 close to $\widetilde{J}_1=1$,  and region 3 with $\widetilde{J}_1>1$. In region 1 and 3, at moderate to high time scales, the value of quantum correlations, both concurrence and discord, have less variation with respect to time and hence tend to attain steady values at large time. However, in region 1, the value of concurrence is lower in comparison to the same in region 3, while quantum discord in region 1 possess a very small finite value, which is much less than that in the region 3, where it can reach the maximum value of approximately 0.2. 
At small values of time, in regions 1 and 3, the quantum correlations show noticeable irregularities in their behavior with respect to $\widetilde{J}_1$. Such irregular values of entanglement, at small times, can be attributed to the non-zero value of the correlator, $t_{xy}$, which eventually vanishes at large time. In region 2, as we vary $\widetilde{J}_1$ from disorder to order phase, the concurrence sharply decreases, becomes minimum at $\widetilde{J}_1=1$ and further increases until it saturates in region 3. For small time scales, in region 2, concurrence has irregular behavior and at moderate time scales, it saturates to a small value of the order of $10^{-2}$. However, quantum discord in this region takes more time to reach the steady value. 
Note that, at large time, the system is always more entangled in the region where the choice of quenched coupling constants are from different phases, as described previously in situation ($ii$), in comparison to when the coupling constants chosen from the same phase, as described previously in situation ($i$).

Finally, a close look in Fig.~\ref{fig:J2_05} shows that in the region 3, bipartite quantum correlations at large time, where we may assume the system tends to reach steady state, is always greater than that of the initial state at time $t=0$. However, in the region 1, such enhancement in the quantum correlations through the evolution process happens only if the coupling strength of the initial state, $\widetilde{J}_1$, is less than 0.5, which is  equal to our choice of the coupling constant of the driving Hamiltonian, $\widetilde{J}_2$. We will follow up this scenario later with further discussions.

\subsection*{Case B: $\widetilde J_2$ in ordered phase}
\label{subsec:ordered}

\begin{figure}
\includegraphics[width=0.3\textwidth]{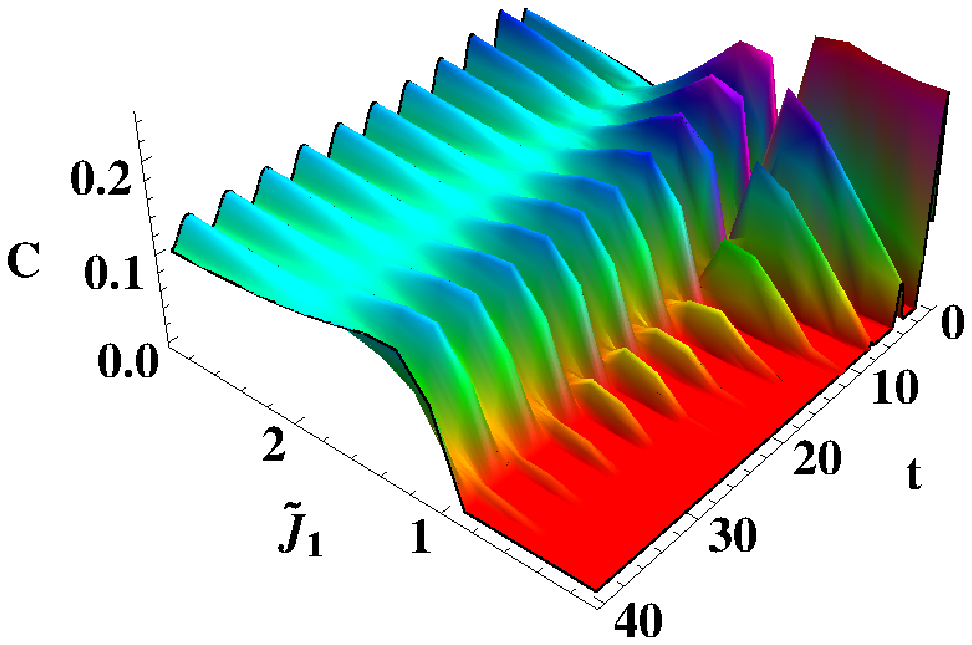}
\includegraphics[width=0.35\textwidth]{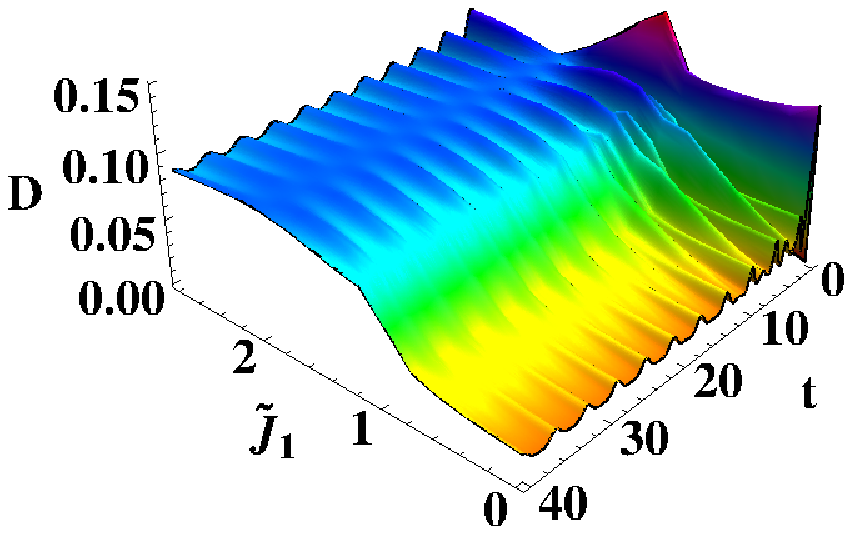}
\caption{(Color online.) Variation of the nearest neighbour concurrence (top panel) and quantum discord (bottom panel) of the evolved state in $XY$ model with respect to quenched coupling constant $\widetilde{J}_1$ and $t$, where$\widetilde{J}_2=2.0$ and $\gamma=0.5$. The dimensions are the same as in Fig. \ref{fig:J2_05}. }
\label{fig:J2_2}
\end{figure}

Let us now fix $\widetilde{J}_2$ of the driving Hamiltonian from the ordered phase. Again the initial state is chosen as the zero temperature state of $H(\widetilde{J}_1)$.

In Fig.~\ref{fig:J2_2}, we plot the concurrence (top panel) and quantum discord (bottom panel) for the system under evolution with respect to the coupling constant corresponding to the initial Hamiltonian $H(\widetilde{J}_1)$ and time. The transverse magnetic field is kept constant, $h=1$, and anisotropic  constant, $\gamma$, is chosen to be as $\gamma=0.5$. We choose $\widetilde{J}_2=2.0$. Similar to the case A, the behavior of quantum correlations can again be analysed by dividing the $J_1$--$t$ plane into three distinct regions: Region 1 with $\widetilde{J}_1<1$, region 2 close to $\widetilde{J}_1=1.0$,  and region 3 with $\widetilde{J}_1>1$. Here we discuss concurrence and quantum discord individually as they have different characteristics.

In region 1, concurrence shows revival and collapse with respect to time. In particular, the amplitude and reviving regions of concurrence gradually decrease and finally vanish at large time.  It is worth mentioning here that the number of revivals that appear in the region 1 depends on the value of anisotropic parameter $\gamma$. As we increase $\gamma$ from zero to one, the number of revivals increases. In the region 3, entanglement oscillates between two non-zero values and the amplitude of oscillations being maximum at low time scales located close to region 2. However, as expected, at $\widetilde{J}_1=2.0$ such oscillations vanish and the entanglement assumes constant non-zero value. In region 2, concurrence shows continuous revival and collapse with the increase in time. The collapse and revival of bipartite entanglement in XY spin chain with time-dependent field have been studied in \cite{collapse_revival}.

At small time scales, irrespective of regions, quantum discord has large irregularities in its strength and as time increases, these irregularities vanish and regular oscillations occur. Comparing entanglement and discord, for at large t,  we find that at $\widetilde{J}_1=1$, entanglement shows much sharper transition from zero to a non-zero value while smooth transition is observed for quantum discord. 
Note that, at large times, the strength of quantum discord is much higher in region 3 than that of region 1, where quantum discord survives with small values. In region 2, quantum discord shows smooth increasing trend from disorder to order phase of $\widetilde{J}_1$.

\subsection*{Dynamical enhancement}
\label{sec:enhancement}

In previous section, we have elaborately discussed the features of quantum correlations measures for two specific cases, Case A and Case B, for two chosen values of the coupling constant, $\widetilde{J}_2$, corresponding to the driving Hamiltonian. We would now like to see whether such features are generic at large times, when the observables acquire steady state values. For this, 
we set $t \to \infty$ in the analytical expressions of the magnetization and the correlators (see appendix A) and continuously vary the coupling constants $\widetilde{J}_1$ and $\widetilde{J}_2$.

\begin{figure}
\includegraphics[width=0.23\textwidth]{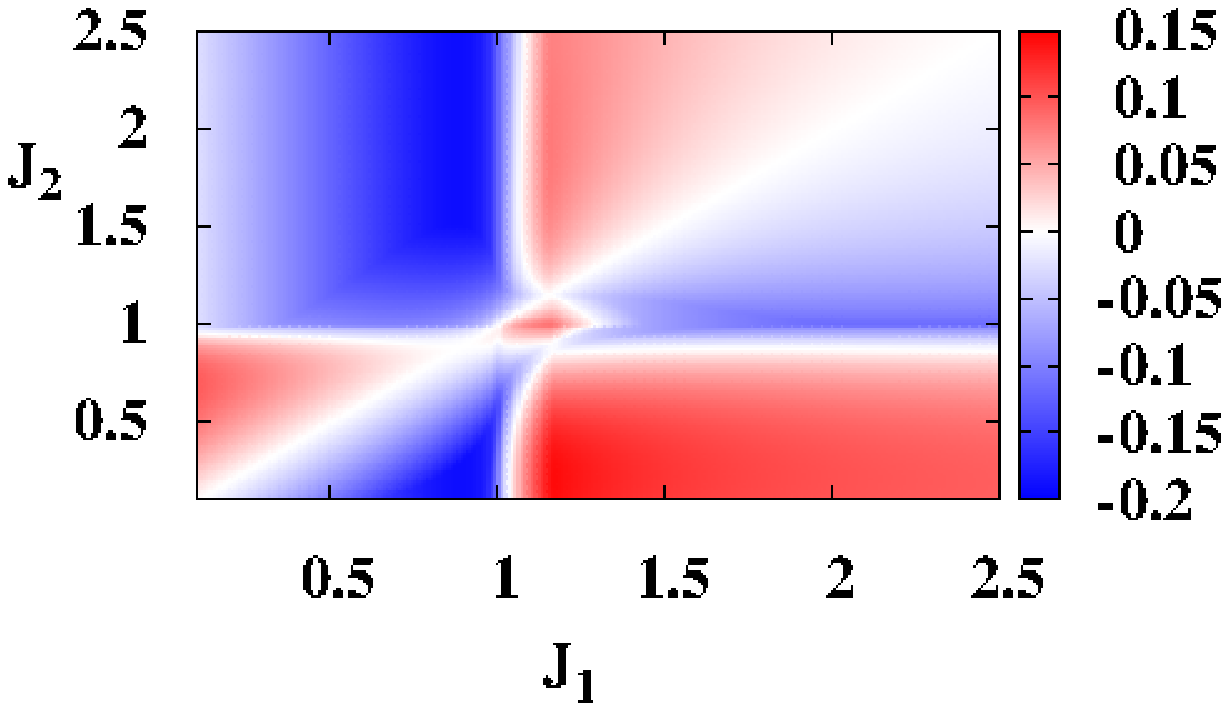}
\includegraphics[width=0.23\textwidth]{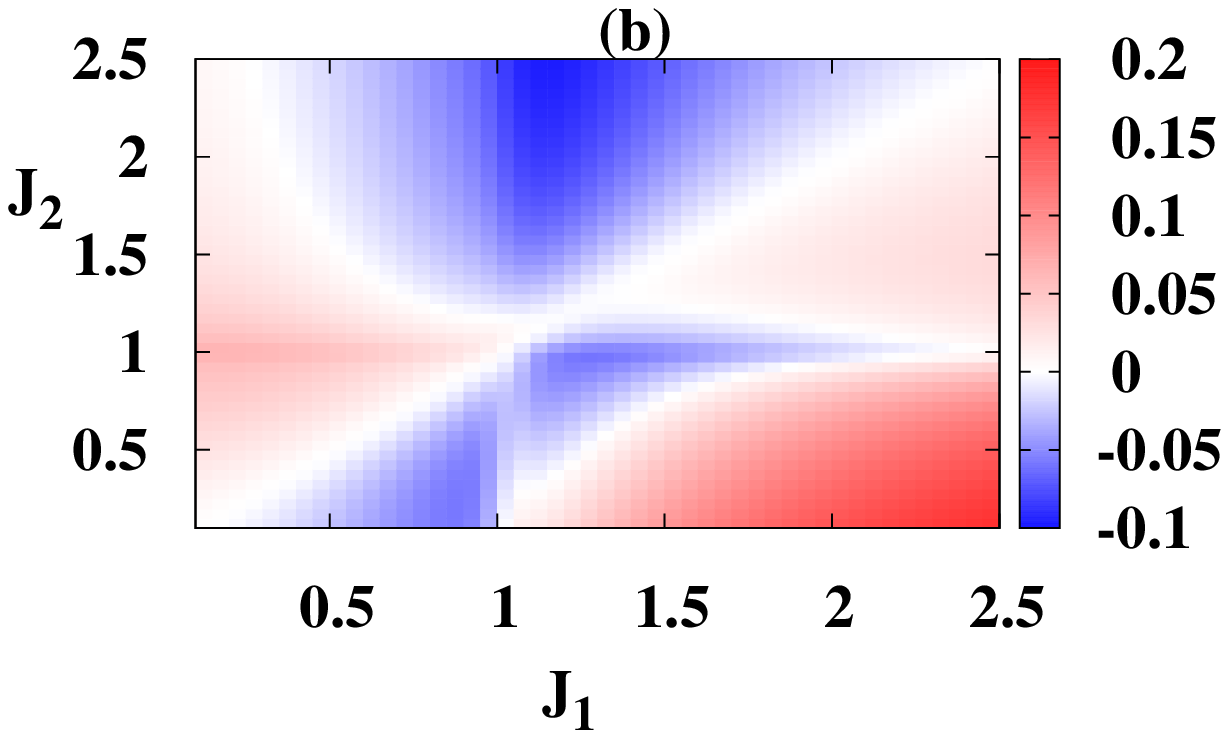}
\caption{(Color online.)(a) Plot of the difference in entanglement between the final and the initial two-site states, $\delta C$, in the quantum $XY$ spin chain against $\widetilde{J}_1$ and $\widetilde{J}_2$. The negative values (the blue regime) in the difference indicates that the initial bipartite entanglement is more as compared to the final bipartite entanglement in the $J_1-J_2$ plane, while the positive value (the red regime) reveals that bipartite entanglement is enhanced during dynamics compared to the initial state.  Here $\gamma=0.5$ and $h=1.0$. (b) The plot of the difference in quantum discord between the final and the initial two-site states, $\delta D$. The dimensions are same as in Fig.~\ref{fig:J2_05}.}
\label{fig:conc-diff}
\end{figure}

In Fig.~\ref{fig:conc-diff}(a), we plot the difference of final and the initial state entanglement, $\delta C$, where $\delta C = C(t\to\infty)-C(t=0)$, as a function of quenched coupling constants $\widetilde{J}_1$ and $\widetilde{J}_2$ for $\gamma=0.5$. Here the external magnetic field is set at unity, and we set $\widetilde{J}_1=J_1$ and $\widetilde{J}_2=J_2$. 

In Fig.~\ref{fig:conc-diff}(a), depending on the ranges of the initial and final coupling constants, we divide the panel into four parametric regimes: (i) region 1: with $J_1$ and $J_2$ both less than unity, (ii) region 2: with $J_1<1$ and $J_2>1$, (iii) region 3: with $J_1>1$ and $J_2>1$, and (iv) region 4: with $J_1>1$ and $J_2<1$. 

\begin{figure}
\includegraphics[width=0.23\textwidth]{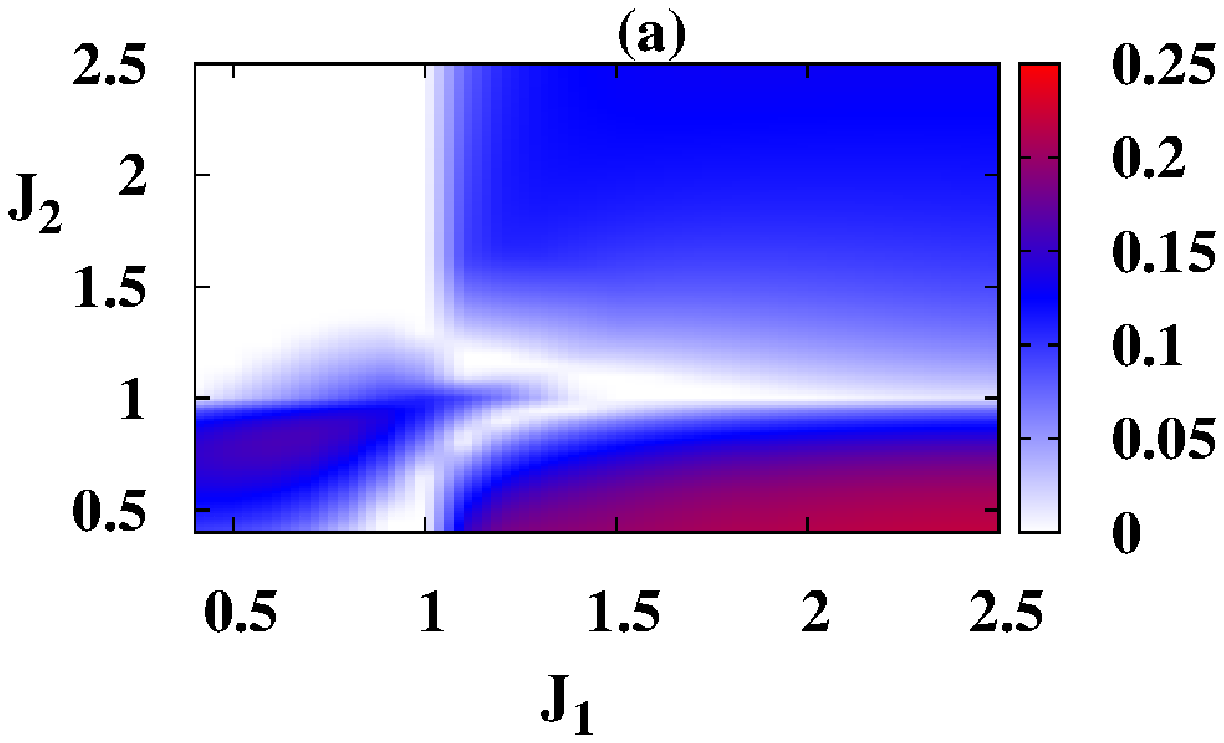}
\includegraphics[width=0.23\textwidth]{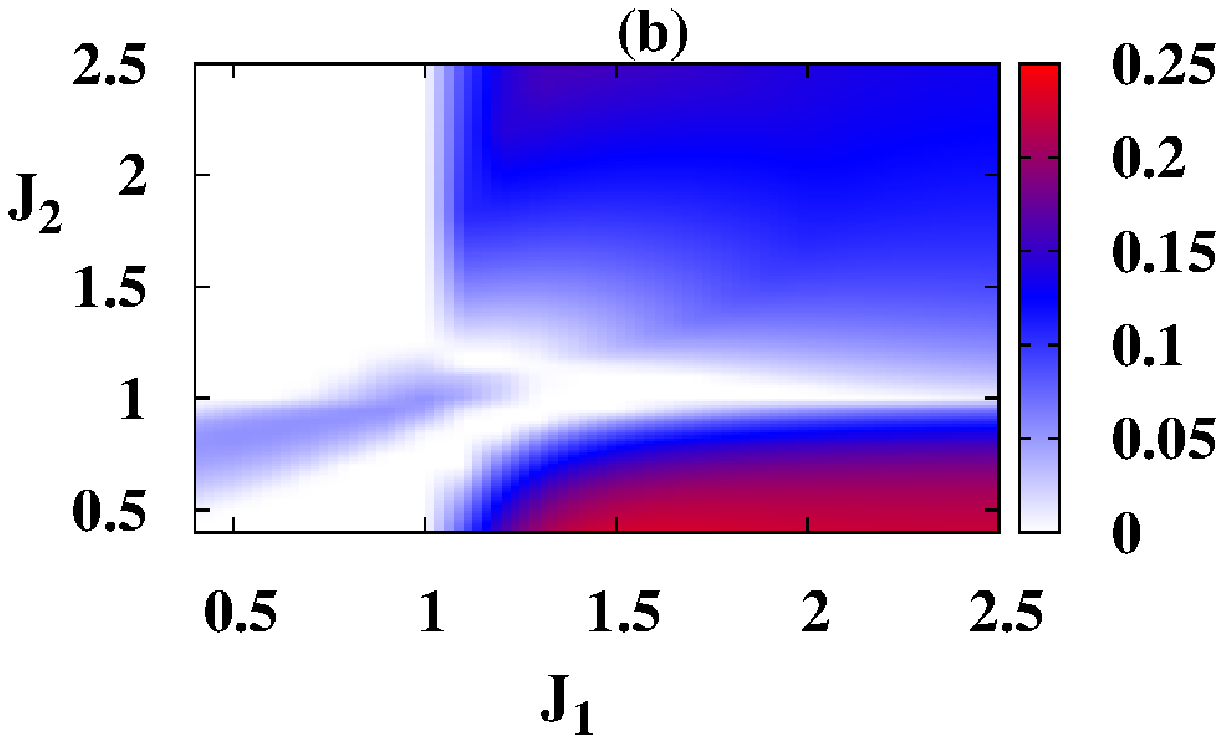}\\
\includegraphics[width=0.23\textwidth]{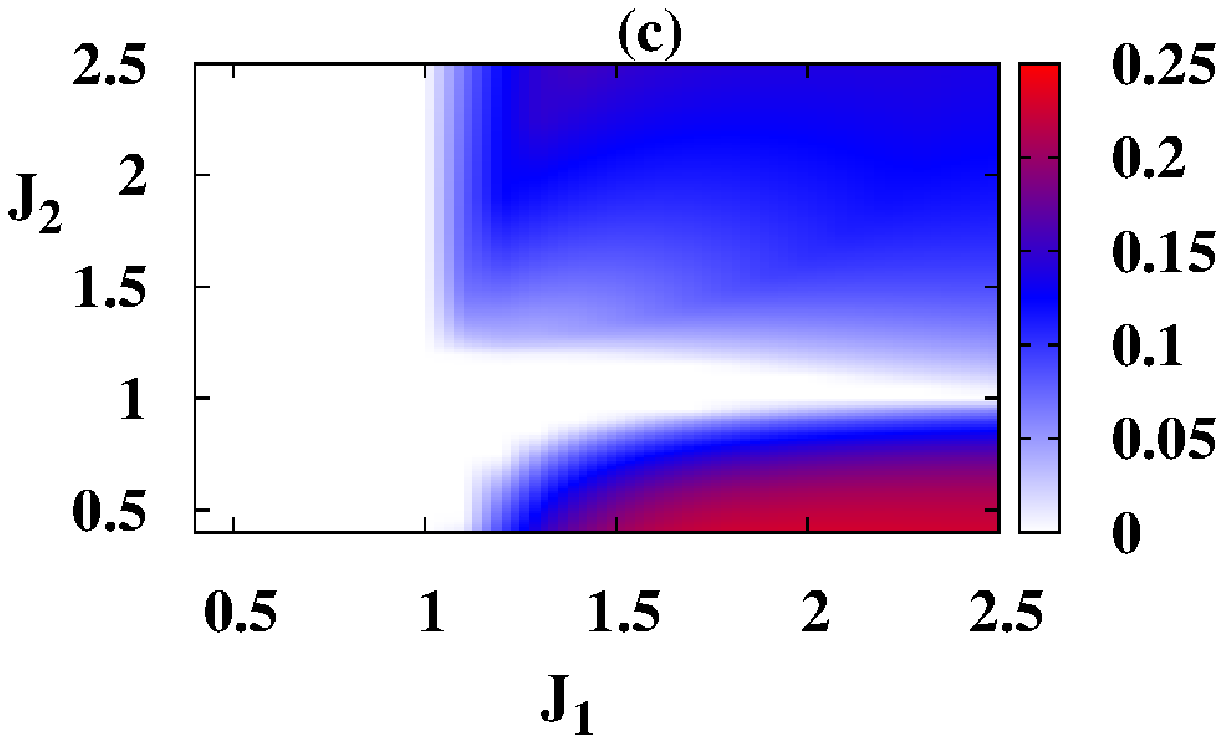}
\includegraphics[width=0.23\textwidth]{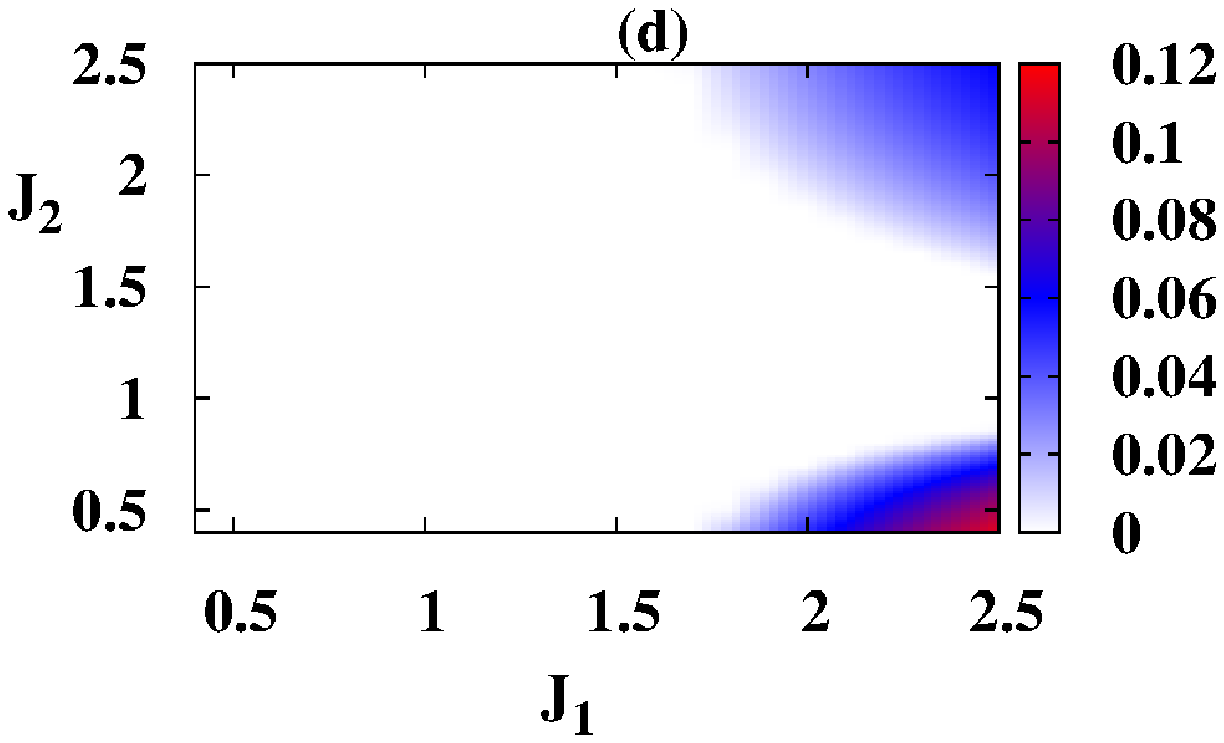}\\
\caption{(Color online.) Long-time behavior of the nearest neighbor concurrence in the quantum XY model with sudden quenching  in the couplings.
The coupling strengths $J_1$ and $J_2$ are varied from $0$ to $3$.  The anisotropy parameter of the  Hamiltonian is fixed at $\gamma=0.5$. The different plots are for different values of $\beta$ of the initial quantum state. (a) $\beta \rightarrow \infty$,  (b) $\beta =3.0$. (c) $\beta=2.0$, and (d) $\beta=0.8$. All the quantities are dimensionless, except concurrence, which is in ebits.}
\label{fig:J1_J2_tinf}
\end{figure}

The white line along the diagonal in Fig.~\ref{fig:conc-diff}(a) corresponds to the case, when $J_1=J_2$. Obviously, in this case the system is not perturbed externally and the entanglement of the system at infinite time is same as in the initial time.  
One can immediately have two interesting observations when the system is quenched across the phase transition point. First, when the initial state is in the ordered phase, i.e., $J_1>1$, and the system is instantaneously quenched to deep in disordered phase, i.e., $J_2 \ll 1$, the nearest-neighbor bipartite entanglement is enhanced significantly than that of the initial state.  For example, for $J_2 = J_1 = 1.4$, the bipartite entanglement of the initial unperturbed state measured by concurrence is approximately $0.066$ ebit, which get enhanced to a value close to $0.186$ ebit by quenching $J_2$ to its final value at $J_2 = 0.2$. 
Secondly, when the initial state is in the disordered phase, i.e., $J_1<1$, and the system is instantaneously quenched to ordered phase with $J_2>1$, the amount of entanglement is significantly decreased compared to the initial value. In fact, any finite entanglement present in the initial state is washed out completely if $J_2$ is chosen from deep ordered phase (see Fig.~\ref{fig:J1_J2_tinf}(a)). As for an example,  the initial state has entanglement 0.143 ebit for $J_2 = J_1 = 0.6$, which vanishes for any driving Hamiltonian  $H(J_2)$, with $J_2>1.3$. 

However, when the system is quenched within the same phase, the steady state bipartite entanglement at infinite time may be enhanced or deteriorated compared to that of the initial state depending on the parametric range. In this situation, we observe that the enhancement occurs when $J_2>J_1$.

Fig.~\ref{fig:conc-diff}(b) shows the difference between the final and the initial state quantum discord, $\delta D$, where $\delta D = D(t\to\infty)-D(t=0)$, as a function of quenched coupling constants $\widetilde{J}_1$ and $\widetilde{J}_2$ for $\gamma=0.5$. The features of quantum discord is approximately similar to the bipartite entanglement. However, unlike entanglement, quantum discord survives with small values when the system is quenched from the disordered to the ordered phase (see also Fig.~\ref{fig:disc_J1_J2_tinf}(a)). We have checked that the qualitative behavior of the entanglement and the quantum discord remain same for other choices of the anisotropy constant, $\gamma$.

\section{Effect of thermal fluctuations}
\label{sec:large_time}

In this section, we consider the thermal state at a given inverse temperature, $\beta$, as the initial state and investigate the effect of temperature of the bipartite quantum correlations of the evolved state at large times.

In Figs.~\ref{fig:J1_J2_tinf}(a-d), we plot the behavior of bipartite entanglement at $t\rightarrow \infty$ as a function of quenched coupling constants $J_1$ and $J_2$ for different temperatures. We choose (a) $\beta=\infty$, (b) $\beta=3$ , (c) $\beta=2$, and (d) $\beta=0.8$. 

To start with, we review the behavior of the entanglement at zero-temperature, but now plot the final state entanglement itself at infinite time as a function of $J_1$ and $J_2$.  We again divide the panel into four parametric regimes as introduced in the previous section.
In Fig.~\ref{fig:J1_J2_tinf} (a), we observe that in region 1 the bipartite entanglement survives with moderate values. In region 2, it is fragile and is close to zero.
In the entire region 3, the entanglement assumes non-zero value. In region 4,  the entanglement is relatively more robust against perturbation, compared to other three regions. We monitor the effect of temperature of the initial thermal state on the entanglement starting from zero-temperature to a finite temperature. It is clear that, in all the regions, the entanglement decreases with the increase of temperature. Comparing Figs.~\ref{fig:J1_J2_tinf}(a)-\ref{fig:J1_J2_tinf}(d), we observe that entanglements in regions 1 and 2 vanish much faster than that of the regions 3 and 4.

\begin{figure}
\includegraphics[width=0.23\textwidth]{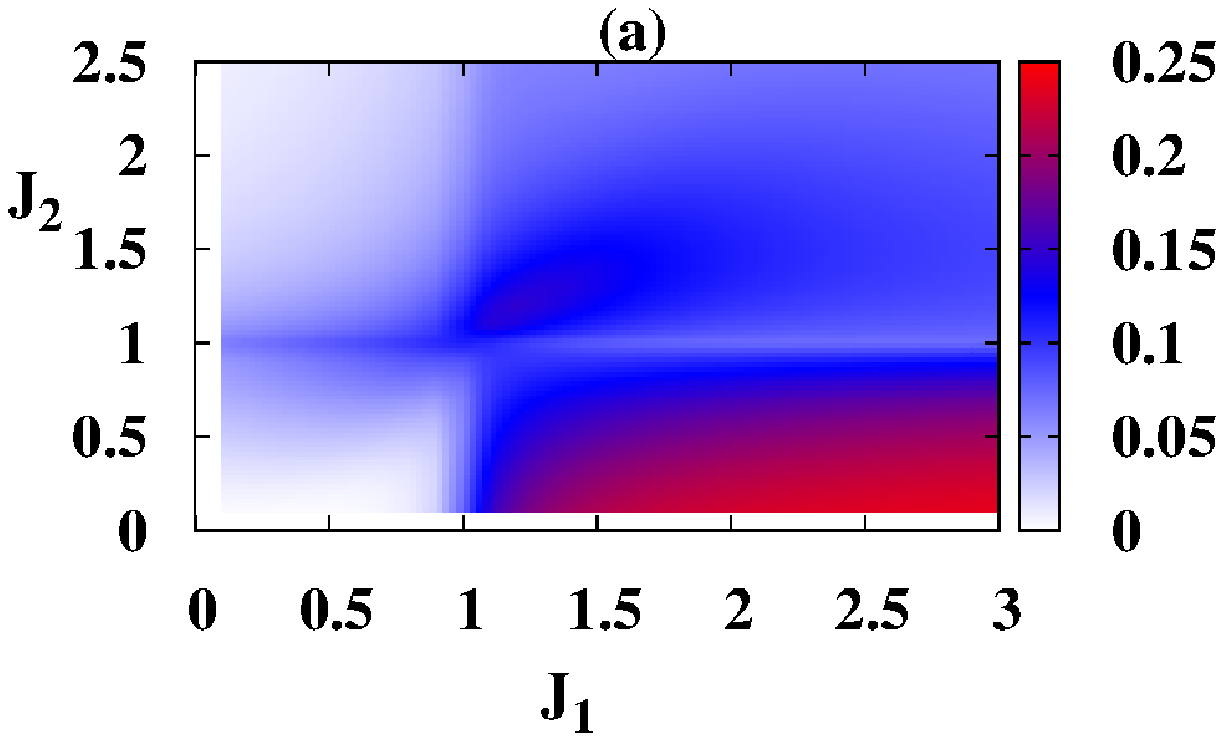}
\includegraphics[width=0.23\textwidth]{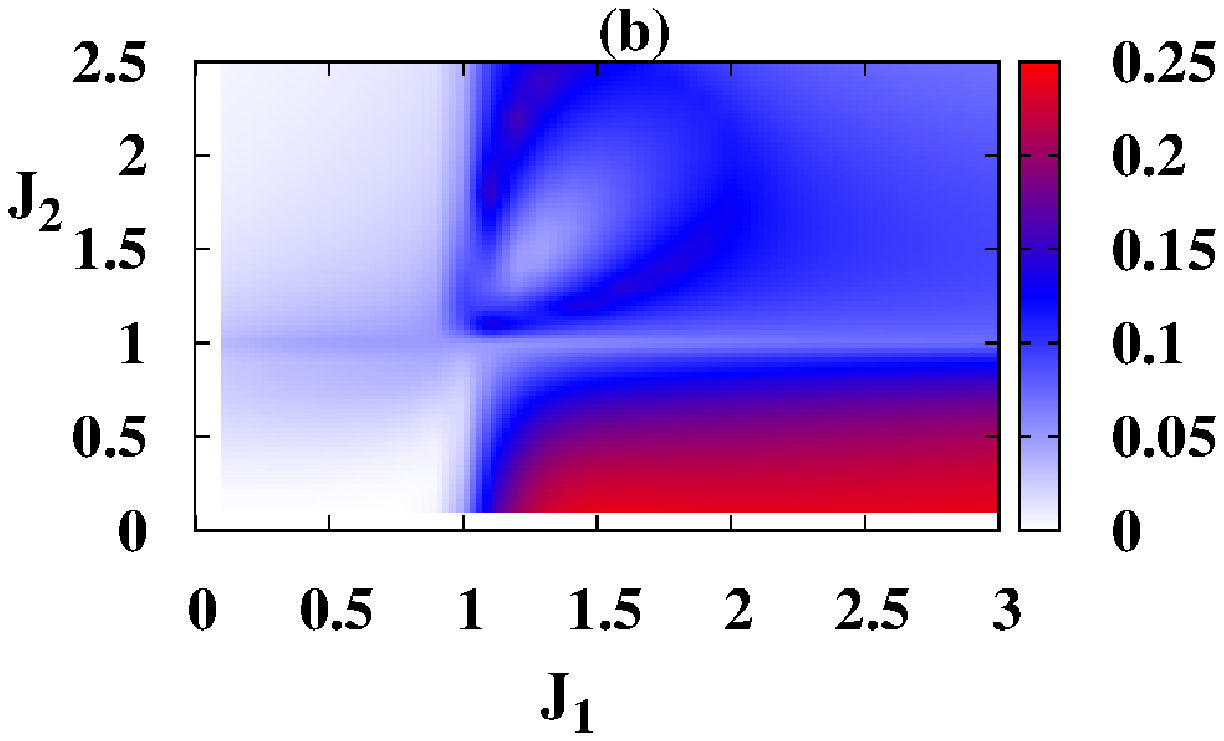}\\
\includegraphics[width=0.23\textwidth]{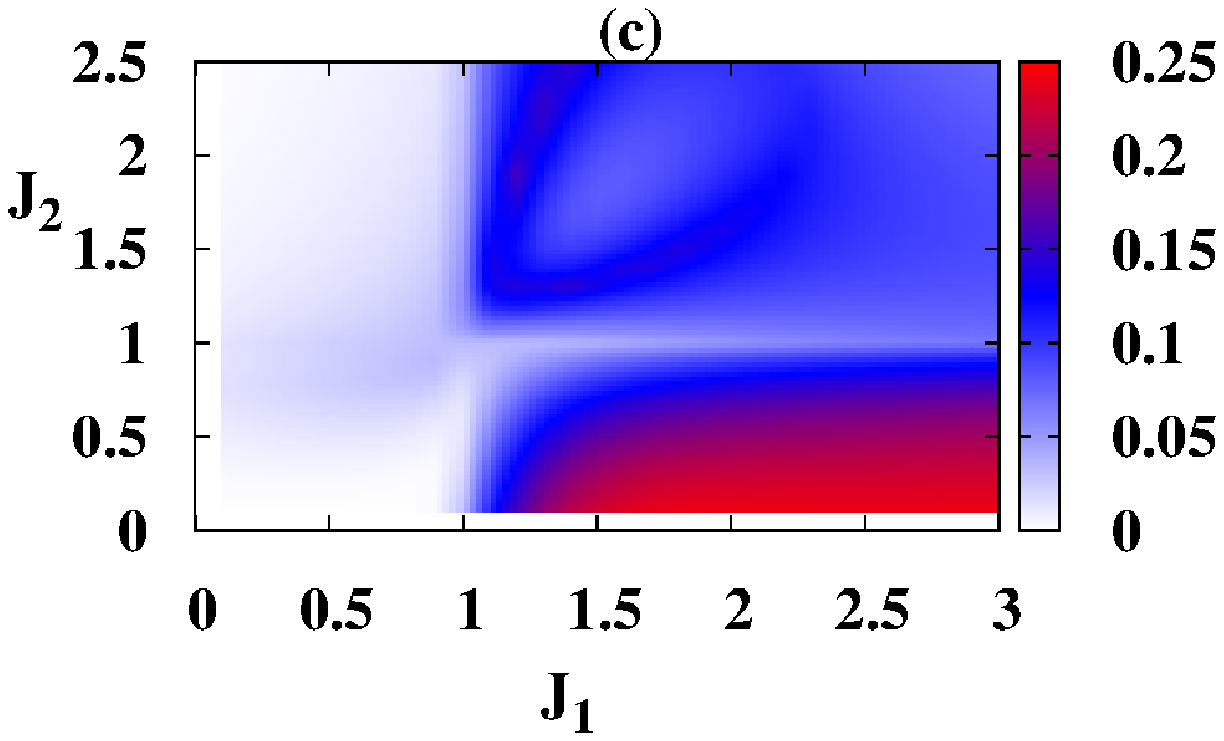}
\includegraphics[width=0.23\textwidth]{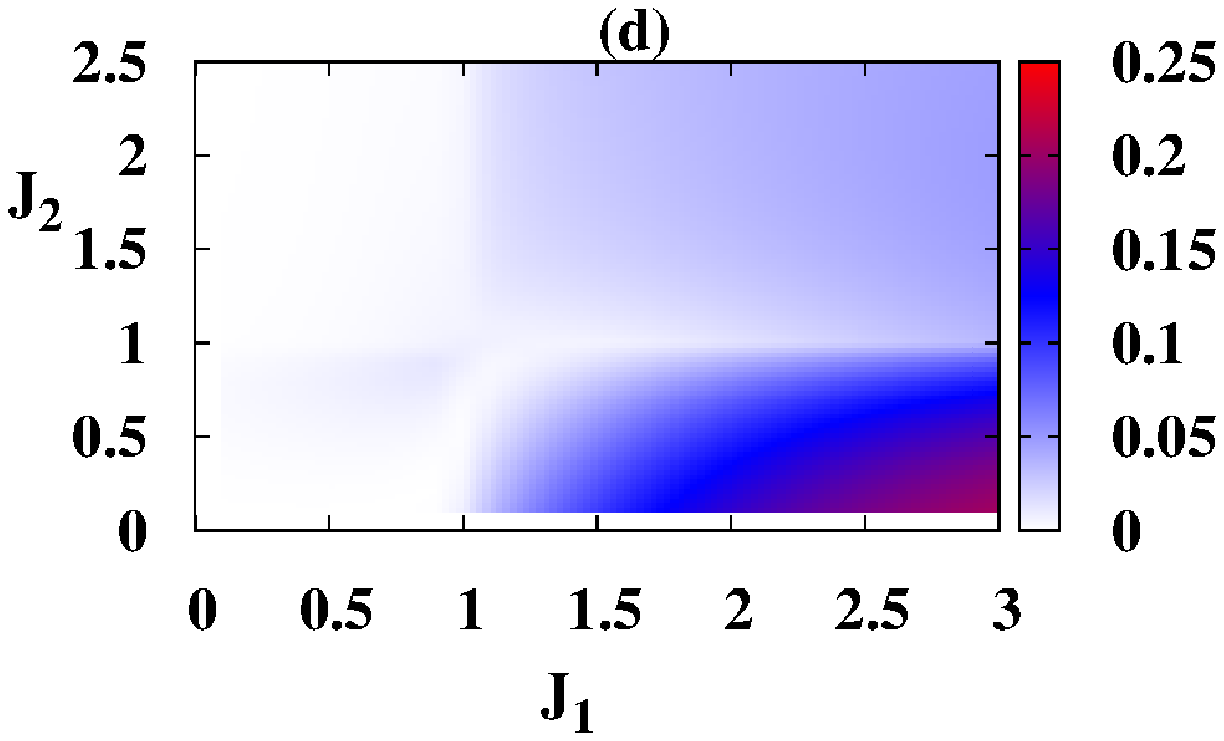}\\
\caption{(Color online.)  Long-time behavior of quantum discord of nearest-neighbour spins in the infinite quantum $XY$ chain with sudden quenching in the couplings. Other details are same as in Fig. \ref{fig:J1_J2_tinf}, except that quantum discord is measured in bits.}
\label{fig:disc_J1_J2_tinf}
\end{figure}

Therefore, we conclude that  the robustness of the entanglement with respect to temperature depends both on $J_1$ as well as $J_2$. 
 The behavior of quantum discord is more or less similar to that of the entanglement, except that the quantum discord is more robust for the increase in temperature as depicted in Figs.~\ref{fig:disc_J1_J2_tinf}(a-d). 
In particular, both quantum correlation measures survives with relatively high temperature when $J_1>1$ and $J_2<1$.

\section{Conclusion}
\label{sec:conc}

Controlled dynamics of isolated complex quantum systems, involving many-particles, in presence of external perturbation has its importance in various fields of theoretical and experimental physics. Quantum correlations, on the other hand, are important resources, for various quantum information and computational tasks. However, they are usually believed to be fragile in presence of the external perturbation. 

In this work, we characterized the dynamics of quantum correlations, such as concurrence and quantum discord, in an infinite XY model between the nearest-neighbor spins due to sudden quenching of the interaction strength,
which is experimentally feasible. 

Two separate cases, where the coupling constant of the driving Hamiltonian, $\widetilde{J}_2$, is chosen either from the ordered phase, i.e., $\widetilde{J}_2>1$ or the disordered phase, i.e., $\widetilde{J}_2<1$, are considered in this paper. With such choices of  $\widetilde{J}_2$, we continuously change the initial state
by varying the initial coupling coupling constant $\widetilde{J}_1$ and temperature. 
Summarizing, for each of the cases, we consider two scenarios of quenching -- either both the initial as well as final coupling strengths are in the same phase or they are across the critical point that connects two different phases. We demonstrate that when the system is quenched from the disordered to ordered phase, any finite entanglement in the initial state eventually vanishes at large times via the dynamical process involving successive collapse and revival phenomena. On the contrary, when the system is quenched from ordered to disordered phase, then the final state blessed with enhanced entanglement compared to the initial state. Such observation can be advantageous in setting up quantum protocols. Since suddenly switching from highly interacting to weakly interacting spin configuration enables us to achieve higher amounts of entanglement between two neighboring spins than that of the initial state.  Also, in the entire parametric regime, the maximum amount of entanglement that can be confined between the two nearest-neighbor spins via quenching exceeds the maximum amount of entanglement that can be present between the two nearest-neighbor spins of the unperturbed system. Moreover, we find that such enhancement of entanglement is also possible via quenching within the same phase, specifically when $\widetilde{J}_2>\widetilde{J}_1$ and  $\widetilde{J}_2$ is not too close to the phase transition point at zero-temperature. We establish that this feature is generic by scanning $\widetilde{J}_2$ itself over the entire range covering the disordered to ordered phases of the driving Hamiltonian and by taking $t \to \infty$, where the system supposedly reaches steady state. We find that the behavior of quantum discord is similar to entanglement. However, unlike entanglement, quantum discord survives with small values when quenched from the disordered to ordered phase.

We extend the analysis from the zero temperature initial state to the initial state with finite temperature, in order to see the effect of thermal
fluctuations on the quantum correlations. We find that irrespective of the quenching scheme, the quantum correlations  are more robust against thermal fluctuation if the initial state is in ordered phase compared to the disordered phase.


\section*{acknowledgement}

We thank Aditi Sen(De), Ujjwal Sen, Himadri Sekhar Dhar and Asutosh Kumar for many fruitful discussions and comments. RP acknowledges an INSPIRE-faculty position at the Harish-Chandra Research Institute from the Department of Science and Technology, Government of India.

\appendix

\section{Magnetization and correlations for infinite $XY$ spin chain}
\label{appendix_sketch}
Following the approach discussed in Refs.~\cite{barouch}, it is possible to obtain exact analytical expressions for the single- and two-body density matrices. Below we briefly discuss the method and present the final expressions.

We define the raising and the lowering spin operators, $b_i^\dagger$ and $b_i$, in terms of the spin operators by $S_i^x= (b_i^\dagger+b_i)/2$, $S_i^y= (b_i^\dagger-b_i)/2 i$ and $S_i^z=b_i^\dagger b_i-1/2$. The raising and the lowering operators are further expressed in terms of Fermi operators $c_j$, where $b_j = \exp\left(-\pi i \sum_{j=1}^{i-1}c_j^\dagger c_j\right) c_i$, and it's complex conjugate, $c_j^{\dagger}$. By performing Fourier series transformation, we obtain new set of operators $a_p$ and $a_p^\dagger$, where $c_j^\dagger = \frac{1}{\sqrt{N}} \sum_{p=-N/2}^{N/2} \exp(i j \phi_p) a_p^\dagger$. Here $\phi_p=2 \pi p/N$.

Expressing the Hamiltonian, $H$, in Eq.~(\ref{eq:HXY}) in terms of the newly introduced operators,  $a_p$ and $a_p^\dagger$, we can write $H = \sum_{p=1}^{N/2} \bar{H}_p$, where
\begin{align}
\label{subspace}
\bar{H}_p=\frac{1}{2}\left[\alpha_p(t) (a_p^\dagger a_p+a_{-p}^\dagger a_{-p}) + i \delta_p(t) (a_p^\dagger a_p^\dagger+a_p a_p)+2 h \right],
\end{align} 
where $\alpha_p(t)=2\left(J(t) \cos \phi_p-h\right)$ and $\delta_p(t)=-2 \gamma J(t) \sin \phi_p$. Recognizing that $[\bar{H}_p,\bar{H}_p']=0$ 
for $p$, $p'=1,2,\cdots,N/2$, the Hilbert space corresponding the Hamiltonian $H$ can be broken down in $N/2$ non-interacting subspaces each of which are in a four-dimensional Hilbert spaces.  Choosing $\{\ket{0}, a_p^\dagger a_{-p}^\dagger \ket{0}, a_p^\dagger \ket{0},a_{-p}^\dagger \ket{0}\}$ as the basis for the $p$th subspace,  $\bar{H}_p$ can be represented in matrix form as 
\begin{equation}
\label{matrixform-H}
 \bar{H_p}= \left( \begin{array}{cccc}
h & \frac {i \delta_p(t)}{2} & 0 & 0\\
- \frac {i \delta_p(t)}{2} & 2 J(t) \cos \phi_p -h & 0 & 0\\
0  & 0 & J(t) \cos \phi_p  & 0\\
0  & 0 & 0 &   J(t) \cos \phi_p \end{array} \right).
\end{equation}
At time $t=0$, we assume the system to be in thermal equilibrium state and the corresponding density matrix for the $p$th
subspace, $\rho_p(0)$ is given by
\begin{equation}
\label{rho-0}
\rho_p(0)=\frac {\exp(-\beta \bar{H}_p)}{\text{Tr} (\exp(-\beta \bar{H}_p))},
\end{equation}
where $\beta=1/({\kappa T})$, $\kappa$ is the Boltzmann constant and $T$ being the temperature of the system. 

Using Eq.~(\ref{matrixform-H}), the matrix form of $\rho_p(0)$ can be obtained as
\begin{equation}
\label{matrixform-rho-0}
\rho_p(0)= \\
 \frac{1}{E(0)}\left( \begin{array}{cccc}
k_{11} & k_{12} & 0 & 0\\
k_{21} & k_{22} & 0 & 0\\
0  & 0 & k_{33}& 0\\
0  & 0 & 0 &  k_{44}  \end{array} \right),
\end{equation}
where 
\begin{widetext}
\begin{eqnarray}
\label{kappas}
E(0)&=&k_{11}+k_{22}+2 \exp\left(-J_1 \beta \cos \phi_p \right),
\nonumber\\
k_{11}&=&\frac{1}{2\Lambda(J_1)}\exp[-\beta\left(-J_1 \cos \phi_p+ \Lambda(J_1)\right)] (\Lambda(J_1)-J_1 \cos \phi_p+h)
+\exp[-\beta\left(-J_1 \cos \phi_p- \Lambda(J_1)\right)] (\Lambda(J_1)+J_1 \cos \phi_p-h),
\nonumber\\
k_{22}&=&\frac{1}{2\Lambda(J_1)}\exp[-\beta\left(-J_1 \cos \phi_p+ \Lambda(J_1)\right)] (\Lambda(J_1)+J_1 \cos \phi_p-h)
+\exp[-\beta\left(-J_1 \cos \phi_p- \Lambda(J_1)\right)] (\Lambda(J_1)-J_1 \cos \phi_p+h),
\nonumber\\
k_{44}&=&\exp\left(- \beta J_1 \cos \phi_p \right)=k_{33},
\nonumber\\
k_{12}&=& i\frac{J_1\gamma \sin\phi_{p}}{\Lambda(J_1)} \exp\left(-\beta J_1 \cos\phi_{p}\right)\sinh(\beta\Lambda(J_1))=k_{21}^{*},~\text{and}~\nonumber\\
\Lambda(J)&=&\sqrt{J^2 \gamma^2 \sin^2 \phi_p+(\cos \phi_p-h)^2}.
\end{eqnarray}
In Eq.~(\ref{kappas}), we have replaced $J(t=0)$ by $J_1$ as assumed in Eq.~(\ref{eq:jt}).
\end{widetext}

Solving the Liouville equation of the system \cite{barouch}, it can be shown that the evolution of the density matrix, $\rho_p(t)$, 
corresponding to the $p$th subspace satisfies 
\begin{equation}
\label{Liouville}
i \frac{d}{dt} \rho_p(t)=\left[H_p(t),\rho_p(t)\right],
\end{equation}
where $p =1, 2, \cdots, N/2$. Considering $U_p(t)$ as the time evolution matrix satisfying 
\begin{equation}
\label{unitary-evolve}
U_p(t)=\exp \left(-i t \bar{H}_p(t)\right), 
\end{equation}
the solution of the Eq.~(\ref{Liouville}) is given by
\begin{equation}
\label{rho-t}
\rho_p(t)=U_p(t)  \rho_p(0) U_p(t)^\dagger.
\end{equation}
Using Eqs.~(\ref{matrixform-H}) and (\ref{unitary-evolve}), we obtain
\begin{equation}
\label{matrixform-U}
U_p(t)= \\
 \exp(-i t J_2 \cos \phi_p)\left( \begin{array}{cccc}
v_{11} & v_{12} & 0 & 0\\
-v_{12}^* & v_{11}^* & 0 & 0\\
0  & 0 & 1 & 0\\
0  & 0 & 0 &  1 \end{array} \right),
\end{equation}
where we put $J(t)=J_2$. $v_{11}$ and $v_{12}$ are given by
\begin{eqnarray}
\label{vs}
v_{11}&=&\frac{i (J_2 \cos \phi_p-h)}{\Lambda(J_2)} \sin\left(\frac{\Lambda(J_2) t}{\hbar}\right)+\cos \left(\frac{\Lambda(J_2) t}{\hbar}\right),
\nonumber\\
v_{12}&=&-\frac{J_2 \gamma \sin \phi_p}{\Lambda(J_2)} \sin\left(\frac{\Lambda(J_2) t}{\hbar}\right).
\end{eqnarray}
Plugging Eqs.~(\ref{matrixform-U}) and (\ref{matrixform-rho-0}) in Eq.~(\ref{rho-t}), we have
\begin{equation}
\label{matrixform-rho-t}
\rho_p(t)= \\
\left( \begin{array}{cccc}
l_{11} & l_{12} & 0 & 0\\
l_{21} & l_{22}^* & 0 & 0\\
0  & 0 & 1 & 0\\
0  & 0 & 0 &  1 \end{array} \right),
\end{equation}
where
\begin{eqnarray}
l_{11}&=&k_{11} |v_{11}|^2+k_{12} v_{11} v_{12}^*+k_{12}^* v_{11}^* v_{12}+k_{22} |v_{12}|^2,
\nonumber\\
l_{12}&=&-k_{11} v_{11} v_{12}+k_{12} |v_{11}|^2-k_{12} v_{12}^2+k_{22} v_{11} v_{12},
\nonumber\\
l_{21}&=&-k_{11} v_{12}^* v_{11}^*-k_{12} (v_{12}^*)^2+k_{12}^*(v_{11}^*)^2+k_{22} v_{11}^* v_{12}^*,
\nonumber\\
l_{22}&=&k_{11} |v_{12}|^2-k_{12} v_{11} v_{12}^*-k_{12}^* v_{11}^* v_{12}+k_{22} |v_{11}|^2.\hspace{2.5em}
\end{eqnarray}

\subsection{Magnetization}
The magnetization operator per spin in $z$-direction is given by $m_z = (1/N) \sum_j \langle \sigma_j^z \rangle$, which can again be written in terms of the $a_p$ and $a_p^\dagger$ operators as $m_z(t) = (2/N) \sum_{p=1}^{N/2} \langle a_p^\dagger a_p + a_{-p}^\dagger a_{-p} -1 \rangle$. Representing $m_p$ in the chosen basis for the $p^{th}$ subspace, $m_z$ can be written as
\begin{equation}
m_z =\frac{2}{N}\sum_{p=1}^{N/2}  \frac{-\left(A_p(0) B_p(t)+4~\text{Re}\left[C_p(0) D_p(t)\right]\right)}{E_p(0)},
\label{mz-1}
\end{equation}
where $A_p(0)=(k_{11}-k_{22})$, $B_p(t)=|v_{11}|^2-|v_{22}|^2$, $A_p(0)=k_{12}$, $D_p(t)=v_{11} v_{12}^*$ and $E_p(0)=\left(k_{11}+k_{22}+2~\exp(-J_1 \beta \cos \phi_p)\right)$.
Using set of Eqs.~(\ref{kappas}) and (\ref{vs}) into the Eq.~(\ref{mz-1}), and simplifying further, the final expression for the magnetization is obtained as
\begin{widetext}
\begin{eqnarray}
m_z&=&\frac{2}{N}\sum_{p=1}^{N/2}  \frac{\tanh \left(\beta \Lambda(J_1)/2\right)}{\Lambda(J_1) \Lambda(J_2)^2} 
\Big[(J_2 \cos \phi_p-h)\left\{(J_1 \cos \phi_p-h)(J_2 \cos \phi_p-h)+J_1 J_2 \gamma^2 \sin^2{\phi_p}\right\}\nonumber\\
& &\hspace{23em} +J_2 (J_1-J_2) \gamma^2 b \sin^2{\phi_p} \cos\left(\frac{2 \Lambda(J_2) t}{\hbar}\right)\Big].
\end{eqnarray}
\end{widetext}

\subsection{Nearest-neighbour correlators}
The nearest-neighbour spin-spin correlators are given by $t^{\alpha \beta}=(1/N) \sum_j \langle s_j^\alpha s_{j+1}^\beta \rangle$, where $\alpha$, $\beta$ stands for $x, y,$ and $z$. The correlators, $t^{xx}$ and $t^{yy}$, are given by $G (R)$, where $R=-1$ and $1$ for the $xx$ and the $yy$ correlator, respectively. Expressing $G (R)$ in terms of the $a_p$ and $a_p^\dagger$ operators, we find $G (R)=\langle T_1 \rangle+\langle T_2\rangle$, where $\langle T_1\rangle =\frac{1}{N} \sum_{1}^{N/2} \left[ 2 \cos(\frac{2 \pi p}{N} R) \langle a_p^\dagger a_p + a_{-p}^\dagger a_{-p} -1\rangle\right]$ and $\langle T_2\rangle=\frac{1}{N} \sum_{1}^{N/2}\left[2 i \sin (\frac{2 \pi p}{N} R ) \langle a_p^\dagger a_{-p}^\dagger + a_{p} a_{-p} \rangle\right]$. Now, the term $\langle a_p^\dagger a_p + a_{-p}^\dagger a_{-p} -1\rangle$, which appears in $\langle T_1\rangle$, has already been calculated while deriving the magnetization, $m_z(t)$. It can be shown that
\begin{equation}
\label{t2-1}
\langle T_2\rangle=\frac{-1}{N} \sum_{p=1}^{N/2}4 \sin \left(\frac{2 \pi p}{N} R\right) \left[\frac{A_p(0)+P_p(t)+i C_p(0) Q_p(t)}{E_p(0)}\right],
\end{equation}
where $P_p(t)=Re[i v_{11}^* v_{12}^*]$ and $Q_p(t)=Re[(v_{11}^*)^2+(v_{12}^*)^2]$. Using Eqs.~(\ref{kappas}), (\ref{vs}) and (\ref{t2-1}), the final expression for $\langle T_2\rangle$ is given by
\begin{widetext}
\begin{eqnarray}
\langle T_2\rangle &=&\frac{1}{N}\sum_{p=1}^{N/2} \frac{2 \sin(\phi_p R) \tanh\left(\beta \Lambda({J_1})/2\right)}{\Lambda(J_1) \Lambda(J_2)^2} (\gamma \sin \phi_p)
\Big[J_2\{(J_2 \cos \phi_p-h)\{(J_1 \cos \phi_p-h)(J_2 \cos \phi_p-h)\nonumber\\
& &\hspace{15em} +J_1 J_2	 \gamma^2 \sin^2\phi_p\}-h (J_1-J_2)(J_2 \cos \phi_p-h) \cos\left(\frac{2 \Lambda(J_2) t}{\hbar}\right)\Big].
\end{eqnarray}
\end{widetext}
In analogous way we find the expression of the the $xy$-correlator, $t^{xy}$, as
\begin{widetext}
\begin{eqnarray}
t_{xy}=\frac{1}{N}\sum_{p=1}^{N/2} \frac {\tanh\left(\beta \Lambda({J_1})/2\right)}{\Lambda(J_1) \Lambda(J_2)} b (J_1-J_2) \gamma \sin^2\phi_p \sin \left(\frac{2 \Lambda(J_2) t}{\hbar}\right).
\nonumber\\
\end{eqnarray}
\end{widetext}
It can readily be seen that $t^{xy}$ vanishes for the equilibrium case, i.e., if $J_1=J_2$. By using Wick's theorem the $zz$-correlator, $t^{zz}$, can be now be expressed as $t_{zz}=m_z^2-G(-1) G(1)+t_{xy}^2$.

\end{document}